\documentclass[letterpaper, 10 pt, conference]{ieeeconf}  

\IEEEoverridecommandlockouts                    
\title{\LARGE \bf
Scenario Based Cost Optimization of Water Distribution Networks Powered by Grid-Connected Photovoltaic Systems
}

\usepackage{subcaption}
\usepackage{tikz}
\usetikzlibrary{tikzmark, positioning, fit, shapes.misc}
\usetikzlibrary{decorations.pathreplacing, calc}
\usepackage{authblk}
\usepackage{xurl}
\usepackage{amsmath} 
\usepackage{amssymb}
\usepackage{mathtools}
\DeclareMathAlphabet{\mathpzc}{T1}{pzc}{m}{it}
\DeclareMathOperator*{\E}{\mathbb{E}}

\DeclareMathOperator*{\argmin}{arg\,min}
\begin{document}

\author{Mirhan Ürkmez} 
\author{Carsten Kallesøe}
\author{Jan Dimon Bendtsen}
\author{John Leth}
\affil{\textit{Aalborg University, Fredrik Bajers Vej 7c, DK-9220 Aalborg,
Denmark}\\ \textit{
(e-mail: \{mu,csk,dimon,jjl\}@es.aau.dk)}}

\maketitle

\thispagestyle{empty}
\pagestyle{empty}


\begin{abstract}
The paper presents a predictive control method for the water distribution networks (WDNs) powered by photovoltaics (PVs) and the electrical grid. This builds on the controller introduced in a previous study and is designed to reduce the economic costs associated with operating the WDN. To account for the uncertainty of the system, the problem is solved in a scenario optimization framework, where multiple scenarios are sampled from the uncertain variables related to PV power production. To accomplish this, a day-ahead PV power prediction method with a stochastic model is employed. The method is tested on a high-fidelity model of a WDN of a Danish town and the results demonstrate a substantial reduction in electrical costs through the integration of PVs, with PVs supplying $66.95\%$ of the required energy. The study also compares the effectiveness of the stochastic optimization method with a deterministic optimization approach.
\end{abstract}

\section{INTRODUCTION}
Water distribution networks (WDNs) transport potable water from its source to the end users. Approximately $7\%-8\%$ of the world's energy is utilized for water production and distribution \cite{WSSEnergy}. Since many countries are trying to increase installed renewable energy sources such as  Photovoltaic (PV) and wind, it is natural to ask whether WDNs can run efficiently with renewable energy sources. Specifically, this paper focuses on controlling water pumps in WDNs powered by grid-connected PV panels  to increase the penetration of renewable energy and reduce energy costs. Pump scheduling is a complex problem because of the nonlinearities governing the network elements and the large size of the networks. The problem is exacerbated by the introduction of PV panels due to the uncertainties in PV power production. Another way to approach the integration of photovoltaics (PVs) and WDNs is to consider it as the design of an Energy Management System (EMS) for a microgrid. Microgrids consist of energy sources, storage devices, and interconnected loads. In this case, the grid can be thought of as having WDNs as its loads and water tanks serving as its storage units.

There has been a significant amount of research focused on optimizing the scheduling of pumps in WDNs to reduce costs. Due to the size of the network and nonlinear pipe equations, it is common to approximate the network with a simpler and sometimes linear model. In \cite{Pour2019EconomicMC}, an Economic Model Predictive Control (EMPC) is employed with a Linear Parameter Varying (LPV) system model. In \cite{Fiedler2020EconomicNP}, the network structure is simplified by dividing nodes into clusters and representing each cluster with a single node. Then a system model is developed from the simplified structure using a Deep Neural Network (DNN) structure. In some works, pipe equations are replaced with linear equations or inequalities to simplify the original problem. This approach is taken in \cite{Baunsgaard2016MPCCO}, where the pipe equations are linearized around an operating point and Model Predictive Control (MPC) is applied, and in \cite{Wang2018EconomicMP}, where EMPC is constructed using a set of linear inequalities derived from relaxing the nonlinear pipe equations. In our study, we assume the presence of an elevated reservoir in the network, which allows us to consider the node pressures to be within an acceptable range. Consequently, we do not need to solve the entire network to verify node pressure constraints. Instead, we adopt a simpler model to represent changes in tank levels. Similarly, a network structure with an elevated reservoir is considered and data is utilized to identify a reduced system model in \cite{Kallese2017PlugandPlayMP}. As there is an elevated reservoir, pressure constraints are not included in problem formulation.

One of the main challenges in integrating photovoltaic (PV) panels with water distribution networks (WDNs) is accurately predicting the power production of the PV panels. To address this challenge, several different estimation models have been developed that use weather measurements, historical power production data, and numerical weather predictions (NWP). NWP has been demonstrated to be effective for making predictions as it can take into account factors such as rain and clouds that can significantly impact PV performance \cite{Zhang2019DataDrivenDP}. NWP data is used to classify days into different categories such as sunny, cloudy, or rainy, which allows for more accurate power production estimates to be made \cite{THEOCHARIDES2020115023,8912017,Mellit2014ShorttermFO}. Some models use physical system properties of PVs to formulate power as a function of weather variables such as radiance, temperature, and wind speed, and then use NWP data to make predictions \cite{HULD20113359,6324407,8810897}. Certain studies have refrained from incorporating NWP due to its limitations. In \cite{Ordiano2017PhotovoltaicPF}, the use of NWP is avoided due to concerns about its availability across all locations. Similarly, in our PV forecasting method,we have chosen not to rely on NWP data due to its limited effectiveness for short-term predictions within a time frame of up to 6 hours \cite{THEOCHARIDES2020115023}. Neural Networks (NN) with different prediction horizons \cite{PPFUS} and multiple inputs types (e.g. NWP, historical production data) \cite{THEOCHARIDES2020115023,Mellit2014ShorttermFO} have also been used in the literature. Lately, long short-term memory (LSTM) networks are being applied because of their success in time series forecasting \cite{PMFP, CHEN20211047, 8912017}. A limitation of most NN-based methods is their inability to quantify prediction uncertainty rendering them unsuitable for integration within a stochastic control framework.

The EMS of microgrids with PV panels connected to main grid has been studied with different applications. In \cite{Parisio2015AnME}, an MPC formulated with mixed-integer linear programming (MILP) is employed as EMS
for multiple residential microgrids powered with PVs. In \cite{Parisio2017CooperativeME}, a distributed cooperative approach is used for a network of microgrids with an aggregator determining the distribution of power and individual microgrids able to deviate slightly from the power profile determined by the aggregator. Since PV power production is highly volatile, some works have used scenario-based optimization \cite{Shen2016AME,Xiang2016RobustEM}. Robust satisfaction of the constraints has also been considered alongside  scenario-optimization in \cite{Luo2018ARO}. 

The main contribution of this paper is a method for the control of the pumps in the WDNs that are powered by grid-connected PV systems. PV power production is modeled with a probabilistic model to account for the uncertainties in the production. A stochastic predictive controller method based on a linear model of the system is used to determine the pump flows. Scenario optimization is used to solve the controller problem by utilizing the probabilistic PV model to sample the uncertain variables. The method is applied to the EPANET, a hydraulic simulation software, model of a medium-sized Danish town’s network (Randers).

The outline of the rest of the paper is as follows. PV power estimation method is given in Section \ref{sec:PvPow}. The model of the network is derived in Section \ref{sec:network}. The control method is explained in Section \ref{sec:control}. The experimental results are presented in Section \ref{sec:application}. The paper is concluded in Section \ref{sec:conc}.
\section{PV Power Forecasting}
\label{sec:PvPow}
This section presents a method for predicting the PV power output, which was first introduced in \cite{9968709}. The method is designed to make day-ahead predictions based on historical power data. At night, the prediction is done in two steps. Firstly, the normalized shape of power output for the following day is estimated, after which a multiplier value for this shape is determined. This multiplier value is then applied to the normalized shape to give the final prediction.

The method used for predicting the shape of the power production data is the Exponentially Weighted Moving Average (EWMA).  Let $\eta \in\mathbb{N}$ denote the current day. At midnight between day $\eta-1$ and $\eta$, the production shape of the current day given the historical data is estimated as
\begin{subequations}
\label{eq:EWMA}
\begin{align}
 X'_{\eta-1} &= \frac{1}{\max(X_{\eta-1})}  X_{\eta-1},
\\
Y_\eta &= \alpha X'_{\eta-1} + (1-\alpha)Y_{\eta-1}
\end{align}
\end{subequations}
where $X_{\eta-1}\in\mathbb{R}^{N_{pv}}$ is the vector of production data for the day $\eta-1$, $N_{pv}$ is the
number of power data points in a day given a sampling time of $\Delta_{pv}$, $X'_{\eta-1}\in [0,1]^{N_{pv}}$ is the normalized daily data, $\alpha\in [0,1]$, and $Y_{\eta}\in [0,1]^{N_{pv}}$ is the moving average normalized daily production of day $\eta$. The parameter $\alpha$ controls the relative weight given to more recent days versus older days.

The next step is to determine the coefficient that will be used to multiply the estimated normalized shape $Y_\eta$ to produce the final prediction. To do this, the optimal multiplier value of the previous days is calculated using the optimization problem
\begin{align}
p_{\tau}=\argmin_{p} \quad  \sum_{i=1}^{N}{(pY_{\tau }^i-X_{\tau }^i)^2}\label{eq:optim1}
\end{align}
where $p\in\mathbb{R}$ is the multiplier, and $Y_{\tau }^i$ (resp. $X_{\tau }^i$) are the $i^{th}$ coordinate of $Y_\tau$ (resp. $X_\tau$). Since the optimal multiplier value $p_\eta$ for day $\eta$ can only be calculated when all power production values $X_\eta$ of the day become available, that is at midnight between day $\eta$ and $\eta+1$, the strategy during the day $\eta$ is to estimate the optimal multiplier $p_\eta$.

To estimate $p_\eta$, the time series of optimal multipliers $p_\tau, \tau=1,\dots,\eta-1$ are  modelled using an Autoregressive Moving Average model ARMA(1,1) given by
\begin{align}
\label{eq:ARMA}
    p_\tau = \mu + \phi p_{\tau-1} +  \theta \epsilon _{\tau-1} + \epsilon _\tau 
\end{align}
where $\epsilon _{\tau}$ and $\epsilon _{\tau-1}$ are the forecast errors at step $\tau, \tau-1$ respectively and $\mu, \phi, \theta$ are the constants whose values are set to minimize the sum of squared errors, $\epsilon_\tau$. The error terms $\epsilon _{\tau}=p_\tau-\hat p_{\tau},~\tau=1,\dots,\eta$ are assumed to be coming from independent zero-mean normal distributions. The estimation of the next value $\hat{p}_\eta$ is made using the calculated optimal value of the previous day $p_{\eta-1}$ as 
\begin{align}\label{eq:phat}
\hat p_{\eta}=\mu + \phi p_{\eta-1}  + \theta \epsilon _{\eta-1}. 
\end{align}
Then, the prediction for the day $\eta$ is made as $\hat p_{\eta}Y_\eta$ at midnight.

The problem of predicting during daytime is handled differently than predicting during nighttime since the production values available during the day can be used to predict future power values in the day. Firstly, sunrise is determined when consecutive power values surpass a predetermined limit. Then, an estimation of the optimal multiplier $p_\eta$ is made at each time step using the power data after the sunrise by solving the problem given by
\begin{align}
     \hat p^m_\eta=\arg \min_{p} \quad   \sum_{i=s_r}^{s_r+m}{(pY_{\eta}^i-X_{\eta}^i)^2}
    \label{eq:least-square}
\end{align}
where $\hat p^m_\eta$ is the estimated optimal multiplier value $m$ time steps after the sunrise time $s_r$ based on the available current day data.

We now have a prior estimation $\hat p_\eta$ and an observation $\hat p^m_\eta$ 
 of the optimal multiplier $p_ \eta$. The prior is assumed to be normally distributed as $ p_{\eta}=\hat p_{\eta}+\epsilon _{\eta}$ where $\epsilon_\eta \sim N(0,\sigma^2)$. Let $z^m_\eta$ denote the random variable associated with the observations given by \eqref{eq:least-square}. We assume that $z^m_\eta$ and  $p_\eta$ are related as $z^m_\eta = p_\eta - \epsilon ^m_\eta $ where $\epsilon ^m_\eta \sim N(0,\sigma_m^2)$. Then, likelihood $\mathcal{L}(p_{\eta} \mid \hat p^m_\eta)=P(z^m_\eta=\hat p^m_\eta  \mid p_{\eta})$ can also be represented with a normal distribution. The variances  $\sigma^2$ and $\sigma_m^2$ are estimated from the sets of available $\epsilon_\tau$ and $\epsilon ^m_\tau, \tau=1,\dots,\eta-1 $ values from the previous days. The observation $z^m_\eta=\hat p^m_\eta$ and the prior estimation $\hat p_\eta$ are combined in a Bayesian setting for a better estimation of $p_\eta$ as  
 \begin{subequations}
 \begin{align}
 \label{eq:bayes}
&P(p_{\eta}=x\mid z^m_\eta=\hat p^m_\eta) \propto P(p_{\eta}=x )P(z^m_\eta=\hat p^m_\eta\mid p_{\eta}=x)\\
\label{eq:bayes2}
& \hat p_{\eta \mid m }=\arg \max_{x} P(p_{\eta}=x\mid z^m_\eta=\hat p^m_\eta)
\end{align}
\end{subequations}
 The posterior probability distribution $P(p_{\eta}|z^m_\eta=\hat p^m_\eta)$ is also normally distributed since both the likelihood and prior are normally distributed. The optimal multiplier estimation $\hat p_{\eta|m}$ is selected as the value which maximizes the posterior probability distribution, as expressed in \eqref{eq:bayes2}.

 Differing from the original paper, we also model the error between the optimal estimation $p_\eta Y_\eta$ and the actual values $X_\eta$ to incorporate it into the calculation of the expected cost value of the controller optimization problem outlined in Section \ref{sec:control}. Let $\delta^i_\tau=X_{\tau }^i-p_\tau Y_{\tau }^i$ denote the error at time $i$ in day $\tau$. The daytime error values $\delta^i_\tau$ are assumed to accept an ARMA(1,0) model given by
 \begin{align}
\label{eq:ARMAERR}
    \delta^i_\tau = \phi^{\delta^i_\tau} \delta^{i-1}_\tau + \epsilon^{\delta^i_\tau} 
\end{align}
where $\epsilon^{\delta^i_\tau}$ are  independent and identically distributed with $N(0,\sigma^{\delta^i_\tau})$. It is also assumed that $\epsilon^{\delta^i_\tau}$ and $p_\tau$ are independent when the value of $X_{\tau }^i$ is not known. Since there is no power production at night, nighttime errors are not modeled. The model is fitted to the time series $\delta^i_\tau, \tau=1,\cdots \eta-1$ and the unknown parameters $\sigma^{\delta^i_\tau},\phi^{\delta^i_\tau}$ have been determined.

\section{Network Model}
\label{sec:network}
A water distribution network typically consists of pipes, pumps, tanks, junction nodes and reservoirs. Water networks are commonly divided into zones sharing similar properties like altitude or water consumption. Water flows through the network, driven by the difference in hydraulic head, which is a measure of the fluid pressure and is equivalent to the height of a fluid held in a static column at a given point. The only dynamic elements in a network are the tanks whose water level change as
\begin{equation}
 A_j\dot{{h}}_j = \sum_{i \in \mathpzc{N}_j} q_{ij}
 \label{eq:tankLevel}
\end{equation}
where $A_j$ is the cross-sectional area, $h_j$ is the level of the tank, $q_{ij}$ is the flow entering the tank $j$, $\mathpzc{N}_j$ denotes the set of neighbor nodes of the tank $j$. Flow through pipes $q_{ij}$  connected to the tanks are nonlinear functions $f_i$ of the demand at each node, tank levels, and the amount of water coming from the pumps. To obtain explicit forms of these functions, demand data for every node $d=[d_1~ d_2\cdots]^T$ must be available; however, this is often not the case. We assume that the total demand of the zones supplied by the pumps can be estimated through available data via time series analysis methods, without needing to know the $d$ vector. Since the $f_i$ functions cannot be found without the $d$ vector, we approximate them using linear models and write out the equations for the tank level change as 
\begin{equation}
 \dot{{h}}(t)= Ah(t) + B_1u(t) + B_2 d_{a}(t)
 \label{eq:reducedModel}
\end{equation}
where $h(t) \in \mathbb{R}^{n}$ includes tank levels, $A \in \mathbb{R}^{n\times n}$, $B_1 \in \mathbb{R}^{n\times m}$, $B_2 \in \mathbb{R}^{n\times 1}$ are constant system matrices and $d_{a}(t)$ is the aggregated demand of controlled zone at time $t$, $u(t) \in \mathbb{R}^{m}$ is the input containing pump flows. 
 \section{Stochastic Predictive Controller}
 \label{sec:control}

This section presents a predictive control method for WDNs that are powered by both PVs and the electrical grid. The method builds on the controller introduced in \cite{IfacPaper} and aims to minimize the economic costs associated with operating the WDN. The problem at time $t$ is formulated as 
\begin{subequations}
\label{eq:MPCForm}
\begin{align}
 &\min_{u_0^t,u_1^t \cdots u_{N(t)-1}^t}  \sum_{j=0}^{N(t)-1} \E[J(h_j^t,u_j^t,t)]
\\
 &h_j^t=A_dh_{j-1}^t+B_{d1}u_{j-1}^t+B_{d2} d_{a}(j-1)
 \label{eq:discState}
 \\
 &h_0^t=h(t)
\\
 &u_j^t  \in \mathcal{U} \subseteq \mathbb{R}^{m}&
 \\
 &h_j^t  \in \mathcal{H} \subseteq \mathbb{R}^{n}& \label{eq:MPCstateCons}
 \\
&h_{N(t)}^t \in \mathcal{H}_{tf} \subseteq \mathbb{R}^{n}
\end{align} 
\end{subequations}
where $\E[J(h_j^t,u_j^t,t)]$ is the expected value of economic cost function $J(h_j^t,u_j^t,t)$, $h^t=[h_1^t\cdots h_{N(t)}^t ] \in \mathbb{R}^{n\times N(t)}$ is the predicted future states, $u_j^t \in \mathbb{R}^{m}$ is the input vector, $N(t)$ is the prediction horizon, $\mathcal{U} \subseteq \mathbb{R}^{m}$ and $\mathcal{H} \subseteq \mathbb{R}^{n}$ denotes the input and state constraints respectively and $\mathcal{H}_{tf} \subseteq \mathbb{R}^{n}$ is the terminal state set. The equation \eqref{eq:discState} represents the discretized version of the continuous system \eqref{eq:reducedModel}. At each time step, which is separated by a time interval of $\Delta_t$, the optimization problem \eqref{eq:MPCForm} is solved, and the first term $u_0^{t}$ of the optimal input sequence $\mathbf{u}^t=[u_0^{t} \cdots u_{N(t)-1}^t] \in \mathbb{R}^{m\times N(t)}$ is applied to the system.

The input constraints for the system are determined by the minimum and the maximum flow rate capacity per unit of time that a pump can deliver. These conditions are expressed as 
\begin{multline}
 \mathcal{U}=
 \{[u_1\cdots u_m]^T \in \mathbb{R}^{m} \mid \forall i : 0\leq u_i \leq \overline u_i\}
 \label{eq:inputConstr}
\end{multline}
where $\overline u_1 \cdots \overline u_m$ are upper flow limits. Tank levels are also constrained so that there is always a sufficient reserve of water on hand in case of emergencies, while also preventing overflow. The set $\mathcal{H}$ can be defined as
\begin{equation}
 \mathcal{H}=\{[h_1\cdots h_n]^T \in \mathbb{R}^{n} \mid \forall i : \tilde h_i\leq h_i \leq \overline h_i \}
 \label{eq:stateConstr}
\end{equation}
The linear model \eqref{eq:reducedModel} does not completely capture the complexity of the entire network, leading to a model-plant mismatch. That means the actual state trajectory $h(t)$ might not remain within the set $\mathcal{H}$ even though the predicted states $h^t$ satisfy state constraints. The problem is dealt with barrier-like exponential functions. First, we rewrite state constraints \eqref{eq:stateConstr} as
\begin{equation}
C_i(h) \leq 0, \quad i=0,1,\cdots 2(n-1)
\label{eq:stateIneq}
\end{equation}
where $C_0(h)=\tilde h_1-h_1$ and the rest of the $C_i$ functions are chosen in a similar manner. The cost function terms are then defined as
\begin{equation}
J_{h_i}(h)=e^{a_i(C_i(h)+b_i)} \quad i=0,1,\cdots, 2(n-1)
\label{eq:barrCost}
\end{equation}
where $a_i,b_i \in \mathbb R_{> 0}$. The parameters $a_i, b_i$ define an area near the boundaries of the state constraints where the cost function $J_{hi}$ becomes significantly high. To avoid these high cost values, the predicted optimal state trajectories $h^t$ will steer clear of this region if possible. This allows the actual states $h(t)$ to remain within the state constraints \eqref{eq:stateConstr} assuming the difference between the predicted state and the actual state is small enough. 

The cost function $J(h_j^t,u_j^t,t)$ includes the cost of electricity purchased from the electrical grid to power the pumps. This electricity is used to supplement the power generated by the PVs, and is only obtained when the PV power is not sufficient to operate the pumps. The electricity purchased from the grid is therefore represented as
\begin{equation}
P_{grid}(t)=\max(0,P_{p}(t)-P_{pv}(t))
\end{equation}
 where $P_{p}$ is the total power used by the pumps and $P_{pv}$ is the power generated by the PV panels and $P_{grid}$ is the power purchased from the grid. As this function is not differentiable at the origin we approximate it with the softplus function which can be written as
 \begin{equation}
sp(x)=\frac{1}{\beta}\log(1+\exp{\beta x})
\end{equation}
where $\beta$ is a constant.
 The power provided to the network by the pump  $i$ is equal to ${q}_{pi}(p^{out}_i-p^{in}_i)$, where ${q}_{pi}$ is the pump flow, $p_{out}^i$ and $p_{in}^i$ are the outlet and inlet pressures of the pump $i$. The inlet pressures $p^{in}=[p^{in}_1~p^{in}_2]^T$ are the pressures of the related reservoirs and are assumed to be constant. The outlet pressures $p^{out}=[p^{out}_1~p^{out}_2]^T$ are given as the output of the linear model
\begin{equation}
p^{out}(t)=C_ph(t)+D_pu(t)
\label{eq:heads}
\end{equation}
where $C_p$ and $D_p$ are found using system identification on data generated by the EPANET model. The power used by the pumps at time $t$ is then equal to $P_{p}(t)=u(t)^T(p^{out}(t)-p^{in}(t))$. The power generated by the PVs is equal to $P_{pv}(t)= X_{\tau}^{i}$ where $\tau=(t-t\bmod T_{day})/T_{day}$ denotes the current day and $i=t/\Delta_{pv}\bmod N_{pv}$ is the time step of the daily power profile $X_{\tau}$ corresponding to time $t$. The overall cost function includes both the electricity expense term and the constraint barrier functions can be expressed as  
\begin{multline}
J(h(t),u(t),t)=\sum_{i=0}^{\frac{\Delta_t}{\Delta_{pv}}-1 }c(t)sp(P_{p}(t)-P_{pv}(t+i\Delta_{pv}))\\+\sum_{i=0}^{2(n-1)}J_{h_i}(h(t))
\label{eq:CostFunc}
\end{multline}
where $c(t)$ is the electricity price. It is assumed that the sampling time for PV data, $\Delta_{pv}$, is shorter than the system sampling time, $\Delta_t$, resulting in multiple PV power values being available within the time period from $t$ to $t+\Delta_t$. The electrical costs are then evaluated individually for each distinct PV power value.

In order to address the control problem outlined in equation \eqref{eq:MPCForm}, we use a scenario-based stochastic programming approach, in which we generate $S \in \mathbb{N}$ number of scenarios. The only stochastic variable appearing in the cost function $J(h(t),u(t),t)$ is the PV production data $X^i_\tau$ which is 0 during nighttime. The values of $X^i_\tau$ during daytime are represented by the equation 
$X^i_\tau = p_\tau Y^i_\tau+\delta^{i}_\tau$, $ \delta^{i}_\tau=\phi^{\delta^i_\tau} \delta^{i-1}_\tau+ \epsilon^{\delta^i_\tau}$ where the values of $Y^i_\tau$ and $\phi^{\delta^i_\tau} \delta^{i-1}_\tau$ have been previously calculated, as explained in Section \ref{sec:PvPow}. We use different sampling processes for problems formulated during nighttime and daytime. When the optimization problem \eqref{eq:MPCForm} is formed at night, the values of $X^i_\tau$ are generated by sampling $p_\tau$ using the equation $\hat p_\tau=p_\tau-\epsilon_\tau$ where $\hat p_\tau$ is the estimated value and $\epsilon_\tau$ is normally distributed with known mean and variance. Additionally, error values $\delta^i_\tau$ are calculated using $\delta^{i}_\tau=\phi^{\delta^i_\tau} \delta^{i-1}_\tau+ \epsilon^{\delta^i_\tau}$, where $\epsilon^{\delta^i_\tau}$ is a normally distributed value with a mean of 0 and a known variance $\sigma^{\delta^i_\tau}$. After sunrise, we start to receive non-zero power production data $X^i_\tau$, which provides us with more information about the variables $\epsilon^{\delta^i_\tau}$ and $p_\tau$. Let the index corresponding to sunrise be denoted as $s$. At time $i_c$ the $\epsilon^{\delta^i_\tau}, i=s+1,s+2 \cdots i_c$ values become deterministic functions of $p_\tau$ given by the equations $\epsilon^{\delta^i_\tau}= g_{i}(p_\tau)=X^i_\tau-p_\tau Y^i_\tau-\phi \delta^{i-1}_\tau$ for $i=s+1,s+2 \cdots i_c$ as the values of $X^i_\tau, i=s+1,s+2 \cdots i_c$ are available at time $i_c$. Given this information, the probability distribution of $p_\tau$ is updated using
\begin{subequations}
\begin{align}
\label{eq:sampleCond}
&P(p_\tau=x \mid \epsilon^{\delta^{s+1}_\tau}=X^{s+1}_\tau-p_\tau Y^{s+1}_\tau-\phi \delta^{s}_\tau,\\
&\cdots, \epsilon^{\delta^{i_c}_\tau}=X^i_\tau-p_\tau Y^{i_c}_\tau-\phi \delta^{i_c-1}_\tau ) \propto P(p_\tau=x,\nonumber\\
&\epsilon^{\delta^{s+1}_\tau}=g_{s+1}(x), 
\cdots,\epsilon^{\delta^{i_c}_\tau}=g_{i_c}(x))\nonumber \\
\label{eq:sampleInd}
& P(p_\tau=x,\epsilon^{\delta^{s+1}_\tau}=X^{s+1}_\tau-x Y^{s+1}_\tau-\phi \delta^{s}_\tau, \\ 
&\cdots \epsilon^{\delta^{i_c}_\tau}=X^i_\tau-x Y^{i_c}_\tau-\phi \delta^{i_c-1}_\tau)=P(p_\tau=x) \nonumber \\
&P(\epsilon^{\delta^{s+1}_\tau}=g_{s+1}(x)) \cdots P(\epsilon^{\delta^{i_c}_\tau}=g_{i_c}(x)) \nonumber
\end{align}
\end{subequations}
The relation \eqref{eq:sampleCond} follows from the conditional probability $P(A \mid B)=P(A, B)/P(B)$. In this case the probability of the given information $P(B)=P(\epsilon^{\delta^{s+1}_\tau}=g_{s+1}(p_\tau), 
\cdots,\epsilon^{\delta^{i_c}_\tau}=g_{i_c}(p_\tau))$ is just a constant number, so \eqref{eq:sampleCond} is expressed as $P(A \mid B) \propto P(A, B)$. The equation \eqref{eq:sampleInd} follows from the independency of $p_\tau$ and $\epsilon^{\delta^{i}_\tau}$ terms. After sunrise, the $p_\tau$ is sampled using the new distribution coming from \eqref{eq:sampleInd} and the error values $\epsilon^{\delta^i_\tau}, i=s+1,s+2 \cdots i_c$ are computed directly as they are functions of $p_\tau$. The remaining values of $\epsilon^{\delta^i_\tau}, i=i_c+1 \cdots $ are sampled as they are sampled at night.

The PV power production, electricity price $c(t)$ and total water demand $d_{a}(t)$ signals can be seen as comprising a periodic signal with a period of 1 day and a relatively minor deviation signal. This is leveraged to increase the chance of finding a solution to the optimization problem \eqref{eq:MPCForm}. The idea is to keep the tank levels at the start of each day within a certain range. If a pumping schedule works for the first day, then it is probable that a similar schedule could be applied to the other days since the initial conditions will be similar. Therefore, the horizon $N(t)$ is chosen to target the beginning of each day and expressed as
\begin{equation}
N(t)=(T_{day}-t\bmod T_{day})/\Delta_t
\end{equation}
where $T_{day}$ is the duration of a whole day. The next step is to figure out which tank levels the trajectories should return to at the start of each day. We define the optimal periodic trajectory of the system as the solution of
\begin{subequations}
\label{eq:optTraj}
\begin{align}
 &(\mathbf{u^*},\mathbf{h^*})=\argmin_{{u_i},{h_i}}  \sum_{i=0}^{(T_{day}/\Delta_t)-1} J({h_i},{u_i})
\\
 &{h_i}=A_d{h_{i-1}}+B_{d1}{u_{i-1}}+B_{d2} d_{a}^*(i-1)
\\
 &{u_i}  \in \mathcal{U} \subseteq \mathbb{R}^{m}&
 \\
 &{h_i}  \in \mathcal{H} \subseteq \mathbb{R}^{n}&
 \\
 &{h_0}=h_{T_{day}/\Delta_t}
  \label{eq:period}
\end{align}
\end{subequations}
where $d_{a}^*$ is the average daily demand profile obtained from the past measurements. The resulting state trajectory $\mathbf{h^*}=[h^*_0 \cdots h^*_{T_{day}/\Delta_t}] \in 
\mathbb{R}^{n\times(T_{day}/\Delta_t+1)}$ is the optimal periodic trajectory  because of the periodicity constraint \eqref{eq:period}. Hence, it is sensible to drive tank levels  close to $h_{T_{day}/\Delta_t}^*$ at the end of each day.  The terminal state constraint set $\mathcal{H}_{tf}$ then can be written as
\begin{equation}
\label{eq:terminal}
 \mathcal{{H}}_{tf}= \mathcal{B}_r( h_{T_{day}/\Delta_t}^*)
\end{equation}
where $\mathcal{B}_r(h_{T_{day}/\Delta_t}^*)$ is the open ball centered at $h_{T_{day}/\Delta_t}^*$ with radius $r$.
 If the problem \eqref{eq:MPCForm} becomes infeasible at any time step $t$, we apply the second term of the input sequence from the previous step $u_1^{t-\Delta_t}$.
\section{Application}
\label{sec:application}
\begin{figure}[h]
\resizebox{0.7\columnwidth}{!}{\input{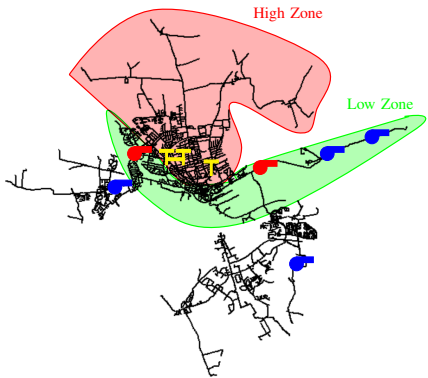}}
\centering
\caption{Water Distribution Network of Randers. The pumping stations to be controlled are shown in red and the remaining in blue. Tanks are shown with a 'T' shaped symbol in yellow.}
\label{fig:network}
\end{figure}

The simulations were carried out utilizing the EPANET software, which is widely used in the industry for creating realistic models and simulating the behavior of networks with a specific pumping strategy. In this particular case, the discussed method is applied to the EPANET implementation of the WDN of Randers, a Danish city, as illustrated in Figure \ref{fig:network}. 
The network contains 4549 nodes and 4905 links connecting them and includes 8 pumping stations. Out of these, 6 are shown in the figure while the remaining 2 are located where tanks are placed. Our goal is to determine the schedules for two pumping stations, while the others operate according to pre-determined strategies. The stations that will be controlled are highlighted in red in the figure, and they provide water mostly to the High Zone (HZ) and Low Zone (LZ). Additionally, there are 3 tanks in the HZ, with two of them connected via pipes and the third standing alone.

 As the pipe connecting the two connected tanks is large enough, water levels at the two connected tanks are almost equal $h_1\thickapprox h_2$ all the time. That enables us to consider $h_1,h_2$ together as one state variable ${h}_{1,2}$.
We have utilized the EPANET model of the network to generate the data needed to approximate the flow of the pipes connected to the tanks as linear functions of the tank levels $h(t)=[{h}_{1,2}(t)~{h}_{3}(t)] \in \mathbb{R}^{2}$, pump flows $u(t)=[{q}_{p1}(t)~ {q}_{p2}(t)] \in \mathbb{R}^{2}$ and the aggregated demand $d_a(t)$. The total demand for the High and Low Zones is used as the aggregated demand in the model since these areas are primarily supplied by the controlled pumps. The model is simulated with various initial tank level conditions and flow rates of the 2 pumping stations that are being controlled, while the control laws for the remaining pumping stations have already been defined in the EPANET model. After fitting the EPANET data to the linear model of the pipe flows, we use the tank level equation \eqref{eq:tankLevel} to obtain the dynamic model of the system. 

\subsection{Simulation Results}
We test our proposed control method on the EPANET model of the Randers water network. Using the Epanet-Matlab toolkit \cite{Eliades2016}, we simulate the network by setting the flow of the two pumps at each time step and controlling the remaining pumps with previously defined rule-based control laws. 

We set the parameters of exponential barrier functions $J_{hi}$ to $a_i=80,$ $b_i=0.3$ for all $i$, and the $\beta$ parameter of the softplus function to $0.02$. It is assumed that the electricity prices are known in advance during the test. 
The maximum tank levels are set to 3m for $h_1$ and $h_2$, and 2.8m for $h_3$, while the minimum tank level is set to half-full. The maximum pump flow is set to 100 and the sampling time $\Delta_t$ to 1 hour, so the control input is recalculated at each hour. The experiments were conducted using a PV power production data set recorded in Albuquerque, US \cite{8980713}. The power values are adjusted by multiplying them with a specific number, in order to align the average power generation throughout the entire data set with the average power consumption of the controlled pumps on a typical day. The average power consumption of the pumps is obtained by simulating the EPANET model for one day. We assume that total demand $d_{a}(t)$ of HZ and LZ can be estimated up to 1 day in advance from the available data. We don't have any actual historical data to measure the demand, so we use a modified version of the actual demand found in the EPANET simulation during MPC calculations.  This modified version is based on a real demand data set from a Danish facility, where the difference between the average demand and the demand on a particular day is added to the EPANET demand. This allows us to replicate an estimated demand. In each experiment a different day from the data set is used, so the assumed estimated demand is different each time.

\begin{figure}
\centering
\begin{subfigure}[b]{0.37\textwidth}
         \centering
         \caption{}{\includegraphics[width=\textwidth]{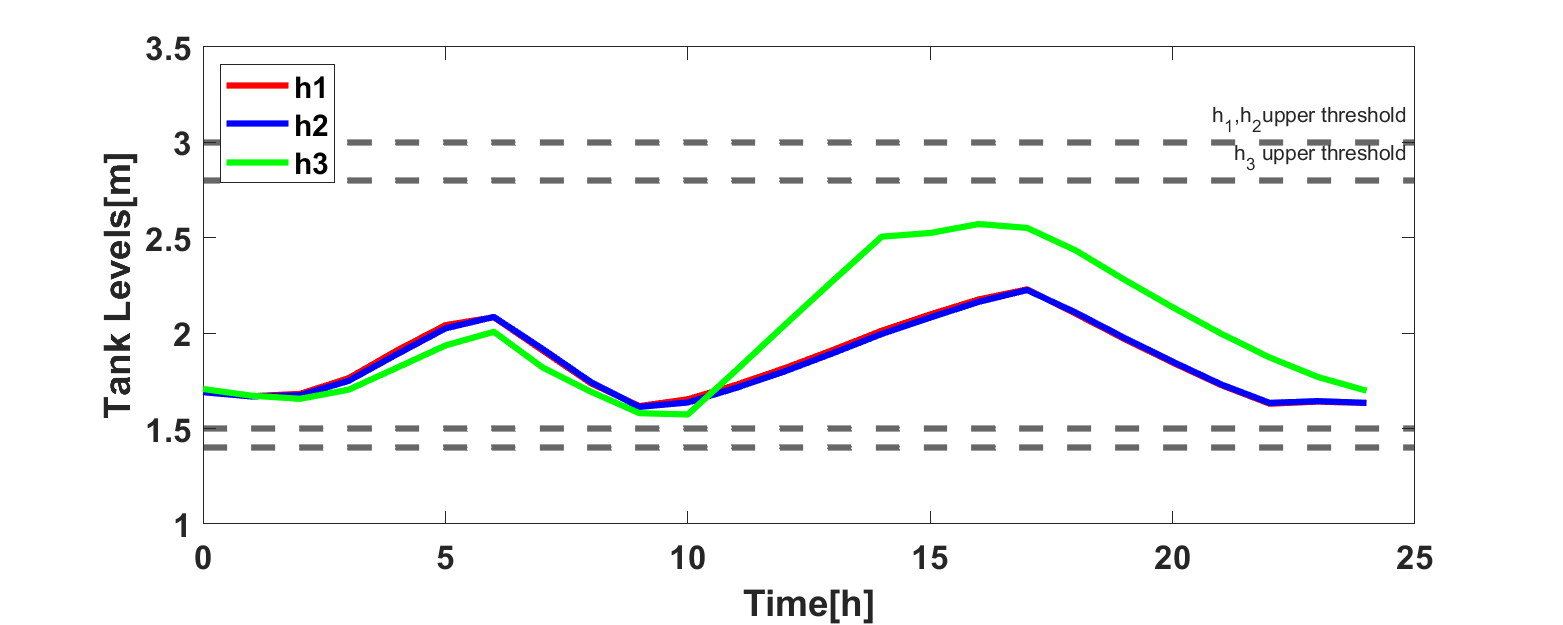}}
\end{subfigure} 
\\
\begin{subfigure}[b]{0.37\textwidth}
         \centering
         \caption{}{\includegraphics[width=\textwidth]{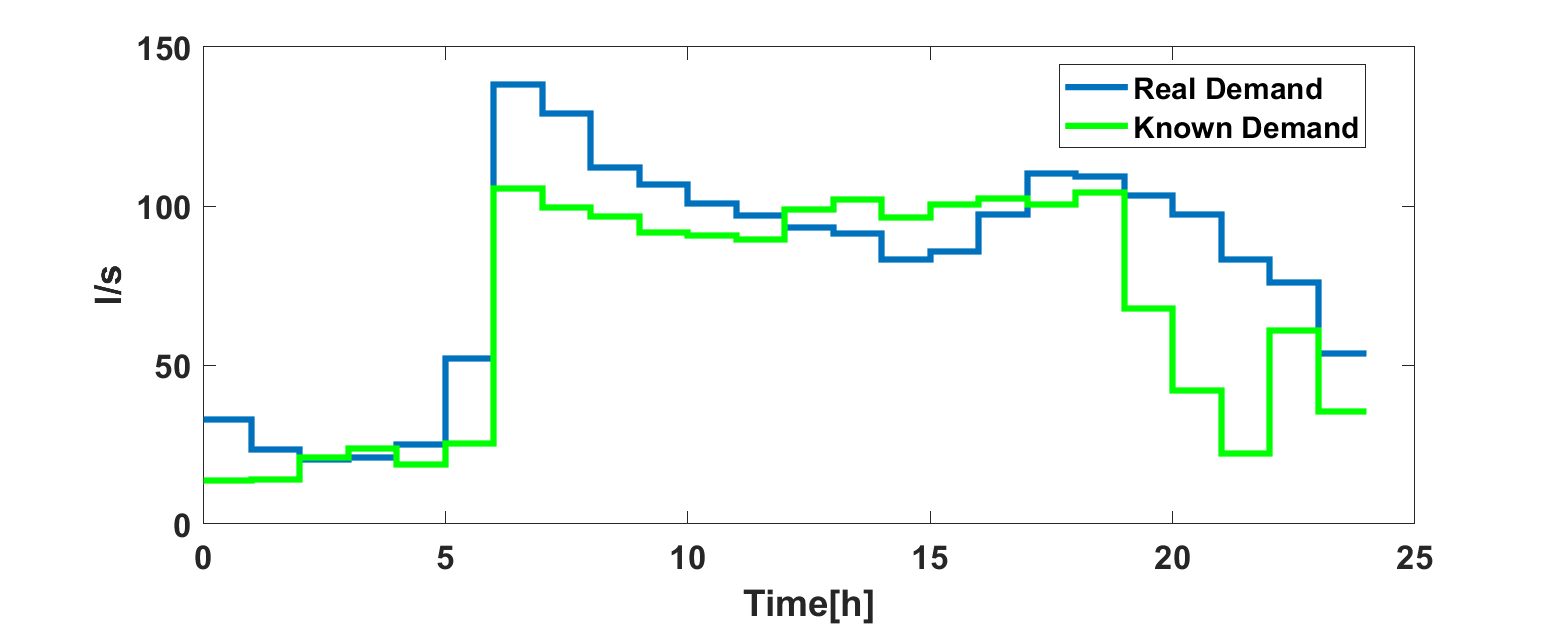}}
\end{subfigure}
\\
\begin{subfigure}[b]{0.37\textwidth}
         \centering
         \caption{}{\includegraphics[width=\textwidth]{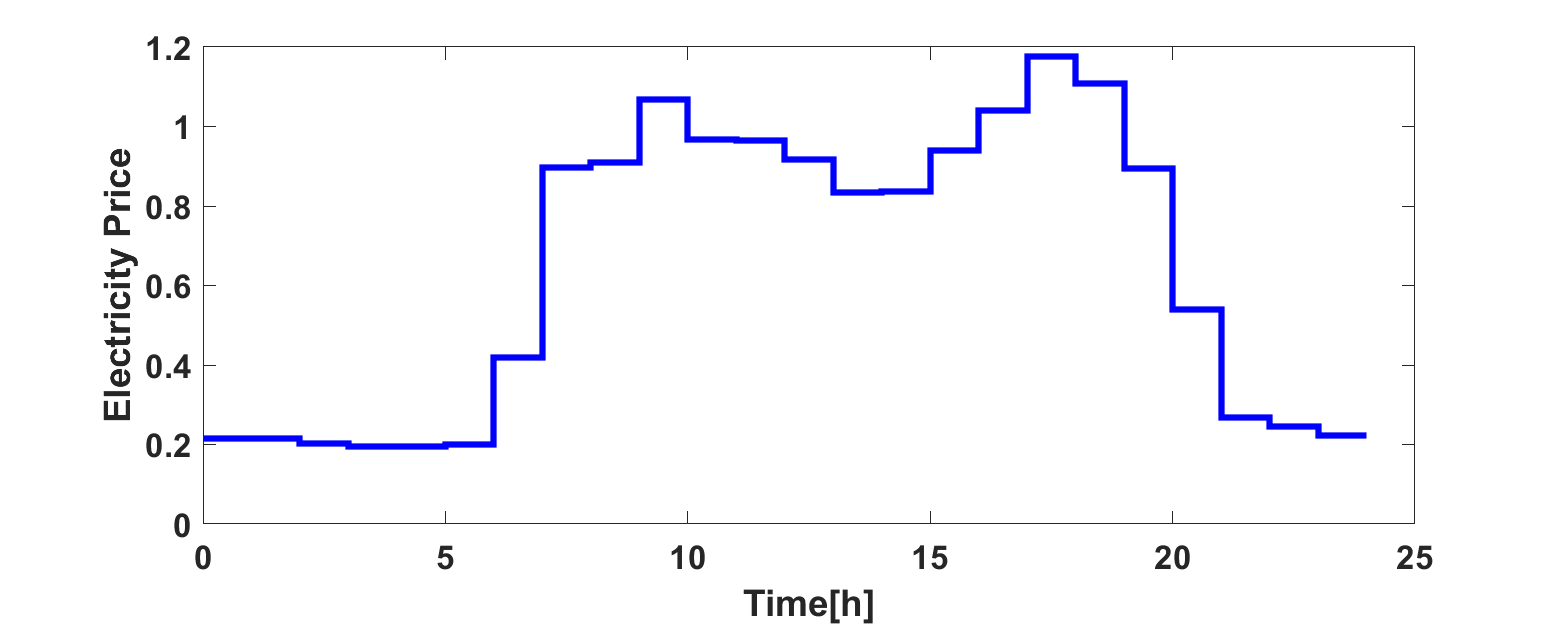}}
\end{subfigure}
\\
\begin{subfigure}[b]{0.37\textwidth}
         \centering
         \caption{}{\includegraphics[width=\textwidth]{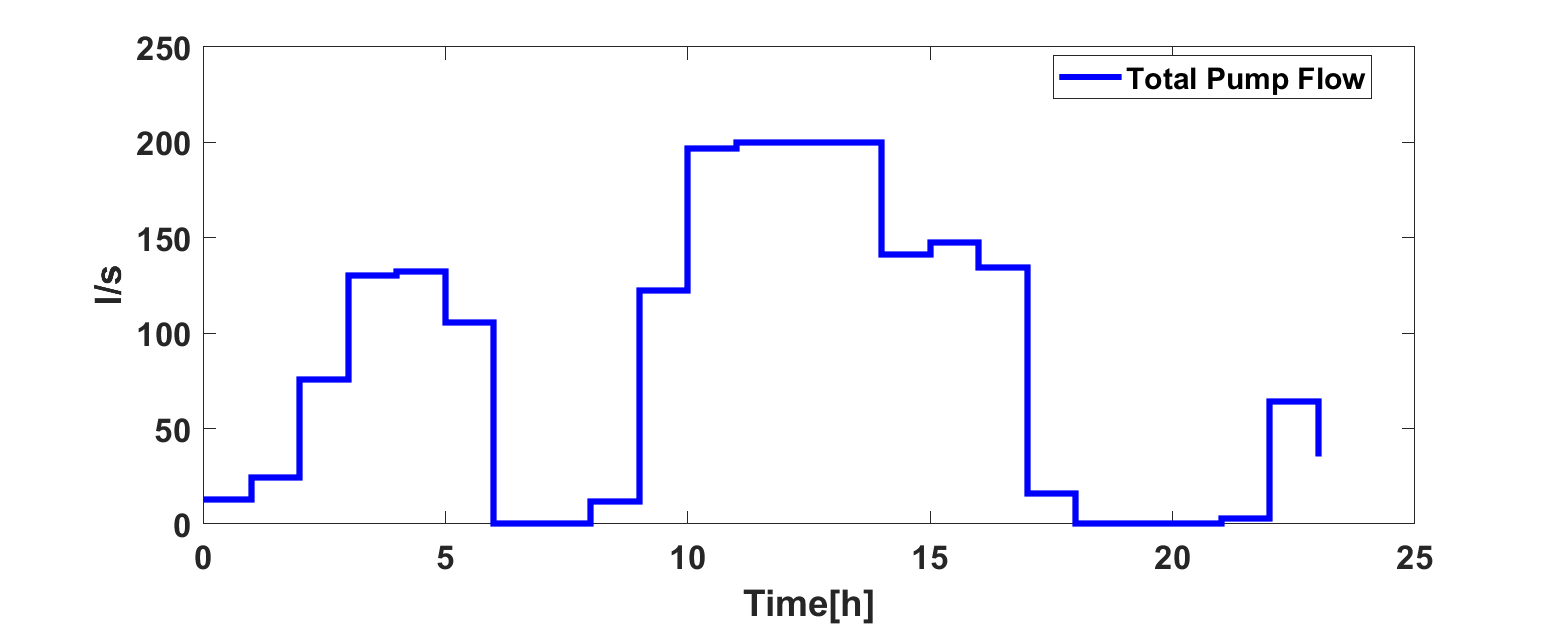}}
\end{subfigure}
\\

\begin{subfigure}[b]{0.37\textwidth}
         \centering
         \caption{}{\includegraphics[width=\textwidth]{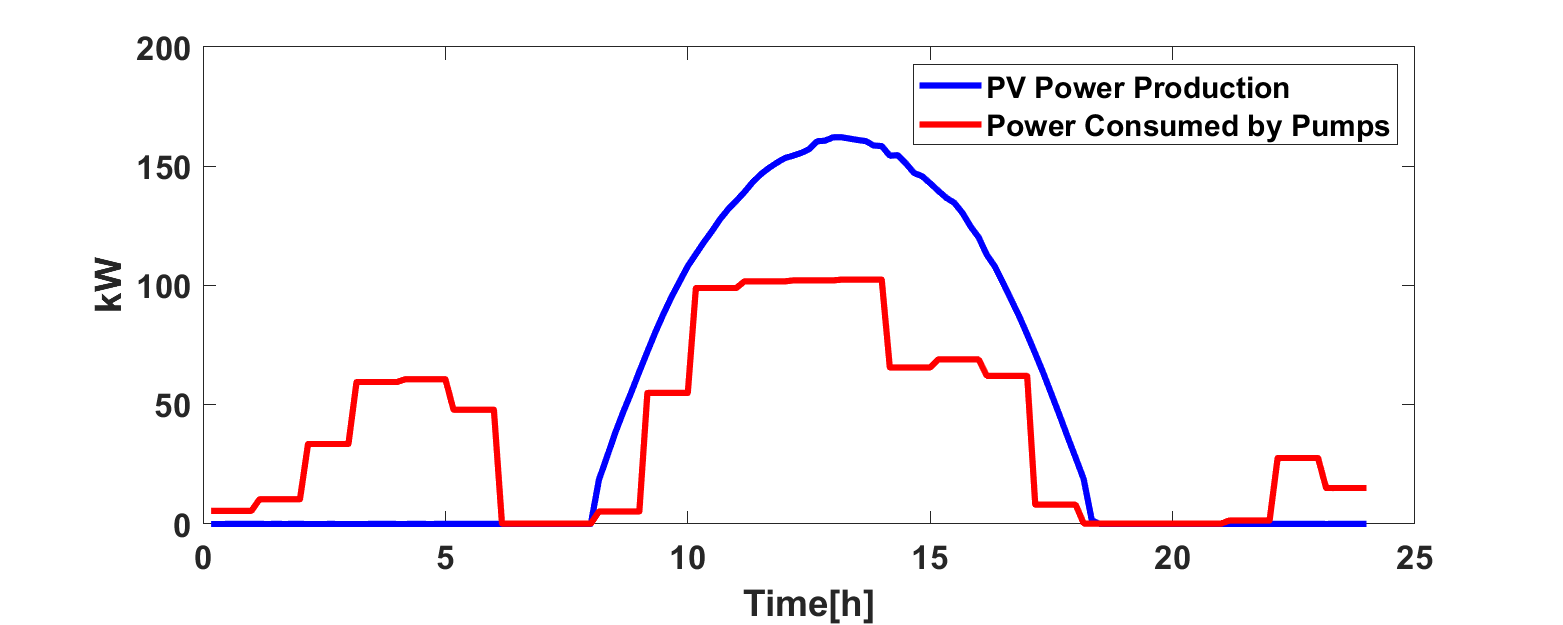}}
\end{subfigure}
    
\caption{Sample simulation. (a) evolution of tank levels through 1 day with upper and lower level thresholds; (b) real total demand of HZ and LZ used in EPANET simulation and the demand used in MPC calculations; (c) total flow provided by the 2 pumps; (d) electricity price; (e) PV power production and the power consumed by the pumps.}
\label{fig:1dayresults}
\end{figure}

The EPANET model of the network is run with the presented method and the results are given in Figure \ref{fig:1dayresults}. The data for PV production is obtained by selecting a single day from the PV data set at random. To make the most efficient use of electricity costs, a moderate amount of water is pumped during the night when electricity prices are low, in order to maintain water levels above the threshold until the PV power can be utilized to pump the water. This strategy can be observed by looking at the water level of $h_3$ around hour 9, where despite the pumping that takes place at night, the water level still approaches close to the lower threshold. To take advantage of the PV energy available during the day, the majority of water is pumped during this period. Once the sun sets and electricity prices decrease again, water pumping continues steadily to ensure that the terminal water level constraints are met. A total of $66.95\%$ of the energy needed for the operation of the pumps is provided by the PV power.

\begin{figure}[bt]
\centering
{\includegraphics[width=0.37\textwidth]{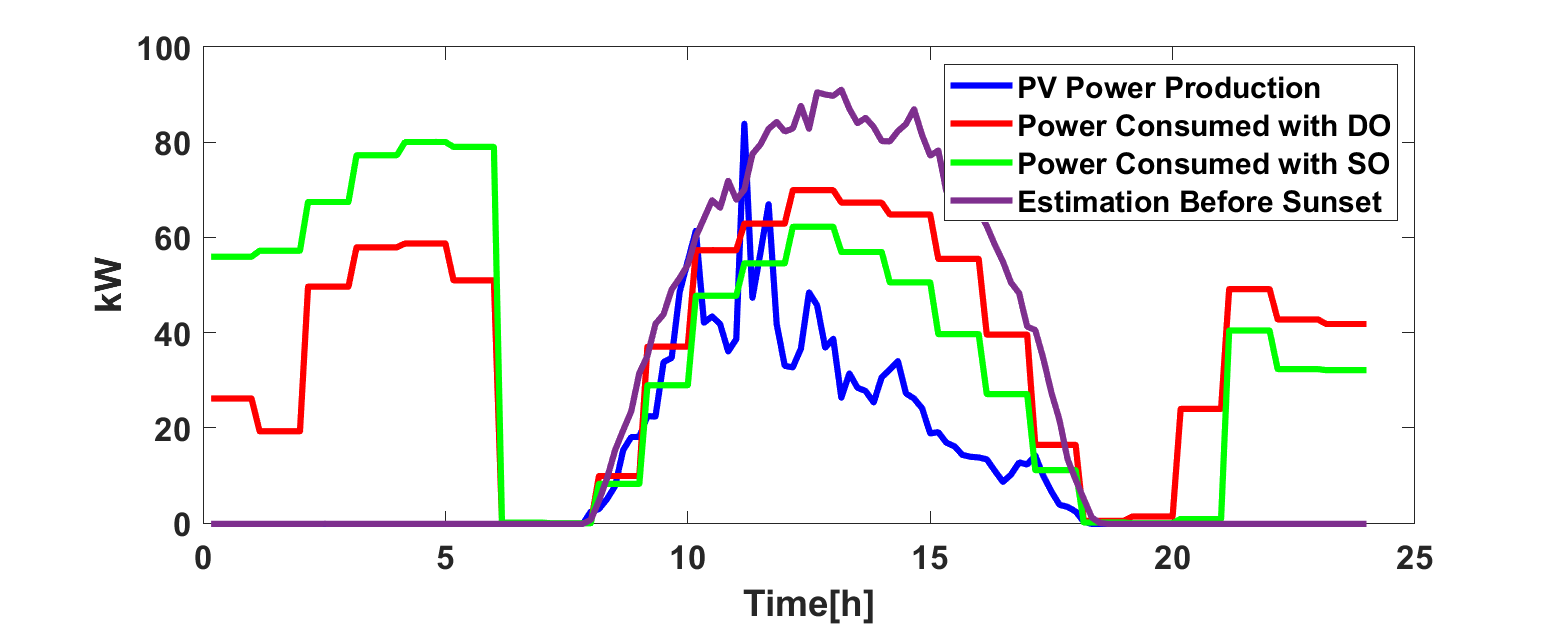}}
\caption{Comparisons of power consumed by pumps with deterministic optimization (DO) and stochastic optimization (SO) along with the power profile and its estimation before the sunset}
\label{fig:badDay}
\end{figure}

The proposed method is then compared with its deterministic optimization counterpart. The key distinction between the two is that the cost function used in the deterministic method is $J(h_j^t,u_j^t,t)$ rather than its expected value $\E[J(h_j^t,u_j^t,t)]$. This means that instead of using a scenario-based optimization, direct estimates of PV power production are used for the calculation of the cost function. To compare the performance of the two methods, the EPANET model is run for 10 days using both methods separately while keeping the demand patterns, PV production profiles, and initial tank levels constant. The PV production profiles used in the experiment are obtained from the dataset by selecting 10 consecutive days. To ensure the robustness and generality of the results, the experiment is repeated 4 times, each with a different randomly selected PV starting day, in order to cover different PV production patterns. The ratios of the total energy costs and the energy used from the grid in the stochastic optimization method with respect to those in the deterministic optimization method are given in Table \ref{tab:relcost} for all 4 cases. In a total of 40 days, electrical costs for the stochastic optimization were $95.92\%$ of the costs for the deterministic optimization. While the costs for stochastic optimization were found to be lower, it should be noted that more energy was drawn from the grid. This is due to the strategy of pumping more water during the night when electricity prices are lower, and less during the day when PV power is available. The reasoning behind this is to avoid purchasing electricity at high prices during times when PV production may fall short of predictions. By using less PV energy and pumping more during the night, stochastic optimization aims to minimize costs. This strategy is particularly beneficial on cloudy days when PV power production is lower than expected. An illustration of the energy consumption for both deterministic and stochastic optimization scenarios on a cloudy day can be seen in Figure \ref{fig:badDay}.  

\begin{table}
\centering
\resizebox{0.7\columnwidth}{!}{
\begin{tabular}{p{6em}|p{6em}|p{8em} } 
 & & \\
 &
  \textbf{Ratio of Electrical Costs} & \textbf{Ratio of Energy Used  From The Grid } \\
 \hline
 Case 1 & 0.9290 & 1.1138 \\
 \hline
 Case 2 & 0.9759 & 1.0682  \\
 \hline
 Case 3 & 0.9734 & 1.1323 \\
 \hline
 Case 4 & 0.9729 & 1.0829  \\
 \hline
 \textbf{Total} & \textbf{0.9592} & \textbf{1.1004}  \\
\hline
\end{tabular}
}
\caption{The ratios of the total energy costs and the energy used from the grid between stochastic optimization (SO) method and deterministic optimization (DO) method $SO/DO$.}
\label{tab:relcost}
\end{table}
\section{Conclusion}
\label{sec:conc}
A method for scheduling the pumps of WDNs powered by grid-connected PVs is proposed to reduce economic costs while taking into account the uncertainties of PV power production through stochastic optimization. Our findings showed that $66.95\%$ of the energy needed for pump operation was supplied by PV power. The integration of PVs resulted in a significant reduction in electrical costs, with the use of stochastic optimization bringing an additional $4\%$ cost reduction compared to deterministic optimization. Stochastic optimization is most effective on days with high uncertainties. However, it does draw more energy from the grid than deterministic optimization, as more water is pumped from the grid when electricity prices are low instead of pumping to the limit when PV power is present.

\bibliographystyle{IEEEtran}
\bibliography{references}
\section*{ACKNOWLEDGMENT}
This work is funded by Independent Research Fund Denmark (DFF). We acknowledge Verdo company, Peter Nordahn, and Steffen Schmidt for providing us with the EPANET model and the network information.
\end{document}